# Quantum Criticality in the Biased Dicke Model

Hanjie Zhu(朱汉杰)[1], Guofeng Zhang(张国锋)[1,2,3,*], Heng Fan(范桁)[4]

[1]*Key Laboratory of Micro-Nano Measurement-Manipulation and Physics (Ministry of Education), School of Physics and Nuclear Energy Engineering; State Key Laboratory of Software Development Environment, Beihang University, Xueyuan Road No. 37, Beijing 100191, China*

[2]*State Key Laboratory of Low-Dimensional Quantum Physics, Tsinghua University, Beijing 100084, China*

[3]*Key Laboratory of Quantum Information, University of Science and Technology of China, Chinese Academy of Sciences, Hefei 230026, China*

[4]*Beijing National Laboratory for Condensed Matter Physics, and Institute of Physics, Chinese Academy of Sciences, Beijing 100190, China*

**The biased Dicke model describes a system of biased two-level atoms coupled to a bosonic field, and is expected to produce new phenomena that are not present in the original Dicke model. In this paper, we study the critical properties of the biased Dicke model in the classical oscillator limits. For the finite-biased case in this limit , We present analytical results demonstrating that the excitation energy does not vanish for arbitrary coupling. This indicates that the second order phase transition is avoided in the biased Dicke model, which contrasts to the original Dicke model. We also analyze the squeezing and the entanglement in the ground state, and find that a finite bias will strongly modify their behaviors in the vicinity of the critical coupling point.**

The Dicke model, which considers a system of a single-mode bosonic field coupled to N two-level atoms, plays a key role as a model illustrating the collective and coherent effects of many atoms in quantum optics [1]. Numerous efforts have been paid to understand its properties over the past few decades, which have resulted in a wide variety of phenomena [2]. One of these interesting phenomena is that the Dicke model undergoes an equilibrium phase transition in the classical limit as the coupling between the atoms and bosonic field reaches a specific value. This type of phase transition is known as the superradiant phase transition (SPT) and has been intensely discussed [3-12].

The early work on the SPT in the Dicke model mostly considers the classical spin (CS) limit, where the number of atoms N tends to infinity. In the classical oscillator (CO) limit, where the ratio of the atomic transition frequency to the bosonic field frequency approaches infinity, the Dicke model also experiences a SPT even at a finite N. This situation had been largely overlooked, and some efforts have been devoted to understand it [10, 12]. This is due to the fact that the Dicke model is mainly realized in the cavity quantum electrodynamics (QED) systems, where the coupling between the atoms and cavity field is weak compared to the atomic and cavity frequencies. Therefore, the achievement of the critical coupling requires using large atomic ensembles. Recently, the superconducting circuit-QED systems have achieved the ultrastrong coupling between the qubit and oscillator [13-18]. This enables us to study the strong coupling effects including the SPT even for a single qubit. The other advantage of the circuit-QED systems is that the qubit parameters can be easily adjusted by the applied bias current, gate voltage, and microwave fields [19]. However, this controllability produces an additional bias term in the Hamiltonian describing the qubit, and thus leads to the biased Dicke model in the circuit-QED systems. While most previous studies have focused on the zero-biased case, one can expect that this additional bias term will produce new phenomena that are not present in the original Dicke model. Several approaches have been developed to describe the behavior of this biased system [20-23]. It has been shown in recent paper that the bias terms will create additional coupling to the environment and increase the fragility of non-classical states [11]. The bias term will also smear out the SPT in the CS limit [22].

The system that we consider is composed of $N$ biased qubits coupled to a single-mode bosonic field. The biased Dicke model describing this system is given by ($\hbar = 1$)

$$H = H_q + H_f + H_{int},  \quad (1)$$

where

$$H_q = \sum_{i=1}^{N}\left(\frac{\Omega}{2}\sigma_z^i + \frac{\varepsilon}{2}\sigma_x^i\right), \quad H_f = \omega a^+ a, \quad H_{int} = \lambda\sum_{i=1}^{N}\sigma_x^i(a + a^+). \tag{2}$$

Here $a$ and $a^+$ denote the annihilation and creation operators for the bosonic field with frequency $\omega$ respectively, and $\{\sigma_i^k; i = x, y, z\}$ are Pauli matrices for the k-th qubit with transition frequency $\Omega$ and bias $\varepsilon$. The field couples to the atoms uniformly with the coupling strength $\lambda$. This system is equivalent to a pseudospin of length $N/2$ coupled to a single mode field, and its Hamiltonian can be written as

$$H = \omega a^+ a + \frac{\Omega}{2}J_z + \frac{\varepsilon}{2}J_x + \lambda J_x(a + a^+), \tag{3}$$

where the angular momentum operators $J_i = \sum_{k=1}^{N}\sigma_i^k, i = x, y, z$.

The biased Dicke model can be realized easily in the system of the flux qubits coupled to a quantum oscillator [18]. In the basis of clockwise and anticlockwise qubit persistent currents, the Hamiltonian of the flux qubit can be written as $H_{qb} = (\Omega\sigma_z + \varepsilon\sigma_x)/2$. Here $\Omega$ is the tunnel splitting and $\varepsilon = 2I_p(\Phi_{ex} - 3\Phi_0/2)$ is the energy bias, where $I_p$ is the persistent current in the qubit, $\Phi_{ex}$ is the external flux threading the qubit loop and $\Phi_0$ is the flux quantum. The interaction between the qubit and the oscillator can be described by $\lambda\sigma_x(a + a^+)$, where $\lambda = MI_pI_0$ is the coupling strength, $I_0 = \sqrt{\omega/(4L)}$ is the measure for zero-point current fluctuations, $L$ is the inductance of the wire. Then the system Hamiltonian reads

$$H = \omega a^+ a + \frac{1}{2}(\Omega\sigma_z + \varepsilon\sigma_x) + \lambda\sigma_x(a + a^+), \tag{4}$$

and is equivalent to the single qubit case of the biased Dicke model.

**Mean-field (MF) theory results**

First we discuss the properties of the system using the mean-field method. Similar to the non-biased spin-boson system [10,24], the mean-field ansatz for the ground-state wavefunction of the biased Dicke model is a product of a spin coherent state $|\theta\rangle$ and a boson coherent state $|\alpha\rangle$, which reads $|\psi_{MF}\rangle = |\theta\rangle \otimes |\alpha\rangle$. The values of the spin inversion and the mean photon number can be written as:

$$\langle J_z \rangle = -N\cos\theta, \quad \langle a^+ a \rangle = \alpha^2. \tag{5}$$

Then the energy functional is given as

$$E(\alpha, \theta) = \omega\alpha^2 + \frac{1}{2}N\varepsilon\sin\theta - \frac{\Omega}{2}N\cos\theta + 2\alpha\lambda N\sin\theta. \quad (6)$$

By minimizing the energy with respect to $\alpha$ and $\theta$ respectively, we acquire the relations

$$\alpha = -\frac{\lambda}{\omega}N\sin\theta, \quad (7)$$

$$-\frac{2\lambda^2}{\omega}N\sin\theta\cos\theta + \frac{\varepsilon}{2}\cos\theta + \frac{\Omega}{2}\sin\theta = 0. \quad (8)$$

The second equation is identical to the semiclassical calculation in Ref. [11]. The Eq. (8) can be expressed as

$$\left(\kappa^2 - \frac{1}{\cos\theta}\right)\sin\theta = \varepsilon'. \quad (9)$$

Here we introduce the rescaled coupling constant $\kappa = 2\lambda\sqrt{N}/\sqrt{\omega\Omega}$ and the rescaled bias $\varepsilon' = \varepsilon/\Omega$. For the $\varepsilon' = 0$ case, our model returns to the original Dicke model, and the mean-field results describes a second-order transition (see Ref. [10]). When $\kappa < \kappa_c = 1$, the solution of the Eqs. (7, 8) is $\alpha = 0$ and $\theta = 0$. Thus the system is in the normal phase. Above the $\kappa_c$ the Eqs. (7, 8) possess nonzero solution and both the field and the atoms acquire macroscopic occupations. This situation corresponds to the superradiant phase.

For $\varepsilon' \neq 0$, this equation can be reduced to a quartic equation, and therefore its solution is not directly applicable to investigate the system. Here we focus on the positive $\varepsilon'$ case, and make some general statements about this solution. When $\kappa \leq 1$, the left hand side of Eq. (9) is a monotonically decreasing function and therefore Eq. (9) only has one negative solution $\theta_-$ [Fig. 1 (c)]. For $\kappa > 1$, the left hand side of Eq. (9) has a local maximum in $\theta > 0$ and a local minimum in $\theta < 0$ [Fig. 1 (b)]. If the coupling constant $\kappa$ exceeds a certain critical value, there can be one negative solution and two positive solutions [Fig. 1 (a)]. In this case, it is easy to verify that the ground state is obtained when using the negative solution $\theta_-$. This negative solution changes continuously with the coupling constant even when $\kappa$ goes across the critical value. For the positive solutions, the larger one is dynamically stable and is corresponded to the high-energy state, while the smaller one is dynamically unstable. Here we call this dynamically stable solution as $\theta_+$.

From the above discussion we can give an intuitive illustration of the effect of the bias term $\varepsilon J_x$. It acts as a symmetry breaking field and creates an asymmetric effective potential. For the positive bias, it lowers the energy functional at $\theta_-$, thus the negative solution is always corresponded to the ground state.

**Quantum corrections to the mean-field theory for the CO limit**

We begin by deriving the effective model for the classical oscillator (CO) limit, where the ratio of the atomic transition frequency to the bosonic field frequency approaches infinity. According to the MF results, for nonzero $\varepsilon$ both the field and the atoms acquire macroscopic occupations, therefore we need to consider the fluctuations above the mean-field ground state $|\theta\rangle \otimes |\alpha\rangle$. To do this, we shift the field operator and rotate the spin operator as follows:

$$a \to a + \alpha; \quad J_z \to J_z\cos\theta + J_x\sin\theta. \tag{10}$$

Here the parameters $\alpha$ and $\theta$ are corresponded to the mean-field state $|\theta\rangle \otimes |\alpha\rangle$, which are the solutions of Eqs. (7, 8). Making these transformations, the Hamiltonian of Dicke model becomes

$$H = \omega a^+ a + \omega\alpha(a^+ + a) + \omega\alpha^2 + \frac{\Omega'}{2}J_z + \lambda\cos\theta J_x(a^+ + a) - \lambda\sin\theta J_z(a^+ + a), \tag{11}$$

where $\Omega' = \Omega\cos\theta - \varepsilon\sin\theta - 4\alpha\lambda\sin\theta$. In this new Hamiltonian, the mean-field ground state is $|-j\rangle \otimes |0\rangle$ thus no macroscopic occupation exists. For the $\Omega/\omega \to \infty$ limit, the low-energy part of the Hilbert space would be confined in the subspace $\mathcal{H}_\downarrow$. By using a unitary transformation $U(\beta, \gamma) = \exp\{iJ_y[\beta(a^+ + a) + \gamma(a^+ + a)^2]\}$, we can decouple the $\mathcal{H}_\downarrow$ and $\mathcal{H}_\uparrow$ subspaces up to second order in $\lambda/\Omega$. Here

$$\beta\Omega' + \lambda\cos\theta = 0, \quad \gamma\Omega' - 2\beta\lambda\sin\theta = 0. \tag{12}$$

Then we can obtain the effective low-energy Hamiltonian by projecting onto $\mathcal{H}_\downarrow$, which reads

$$H_{CO} = \langle -j|U^+ H U|-j\rangle$$
$$= \omega a^+ a + N\beta\lambda\cos\theta(a^+ + a)^2 + \omega\alpha^2 - N\frac{\Omega'}{2}. \tag{13}$$

With squeeze operator $S(\xi)$, this Hamiltonian can be easily diagonalized to give $H'_{CO} = \omega' a^+ a + C$. Here $C$ is a constant we do not care about, and

$$\xi = \frac{1}{4}\ln\left(1 - \frac{4N\lambda^2\cos^2\theta}{\omega\Omega'}\right), \quad \omega' = \omega\sqrt{1 - \frac{4N\lambda^2\cos^2\theta}{\omega\Omega'}}. \quad (14)$$

Now we have obtained the excitation energy $\omega'$. For $\varepsilon' = 0$, the excitation energy is found to be $\omega' = \omega\sqrt{1 - \kappa^2}$ and vanishes at $\kappa = 1$. This vanishing energy scale locates the QPT in the original Dicke model [10]. For $\varepsilon' > 0$, the low-energy properties are very different from the $\varepsilon' = 0$ case. When the coupling constant $\kappa$ is below the critical value, the Eq. (9) only has one negative solution $\theta_-$, and this negative solution gives the low-energy sector of $H$. Above the critical coupling value, the Eq. (9) produces three solutions: one of them is an unstable stationary point, while the other two solutions $\theta_-$ and $\theta_+$ lead to two spectrums. For finite value of $\varepsilon'$, these two spectrums are not identical. From the numerical results we know that $\Omega'(\theta_-) > \Omega'(\theta_+)$ for positive $\varepsilon'$, therefore the spectrum which corresponds to the positive solution is being lifted, while the other spectrum is being lower. This can be shown in Fig. (2). In this figure we plot the lowest 30 energy levels compare to the ground state as a function of the coupling constant $\kappa$. Below the critical point, the low-lying energy levels are similar to that of a harmonic oscillator. Above the critical point, the energy spectrum turns into two equally spaced sets. The energy gap between these two sets is of the $O(\varepsilon)$. For the $\Omega/\omega \to \infty$ limit, this gap is much larger than the excitation energy $\omega'$. Therefore we can reasonably regard the lower one as the low-energy spectrum.

It is obvious that the rescaled excitation energy $\omega'/\omega$ only depends on the rescaled coupling constant $\kappa$ and the rescaled bias $\varepsilon'$. In Fig. 3 we show the convergence to this analytical value when the CO limit is approached. This confirms that $\omega'$ is the real excitation energy in the CO limit.

Crucially, we note that the excitation energy remains finite for arbitrary $\kappa$ in the presents of finite bias $\varepsilon'$. According to the Eq. (14), the excitation energy vanishes only if $\tan\theta = \sqrt[3]{\varepsilon'}$, and can be never fulfilled for the negative solution. We show in Fig. 4 the rescaled excitation energy as a function of the rescaled coupling constant $\kappa$ and the rescaled bias $\varepsilon'$. As in this figure we see that the excitation energy becomes zero only at $\varepsilon' = 0$ and $\kappa = 1$, which indicates the QPT. This scenario is strongly suppressed as $\varepsilon'$ is increased, and there is no longer any sign of a critical point. Thus we conclude that the characteristic energy scale of fluctuations above the ground state does not vanish for arbitrary coupling, and the second order phase transition cannot occur for finite bias.

For the positive solution, the relation $\tan\theta = \sqrt[3]{\varepsilon'}$ can be satisfied when the coupling reaches the critical value, and it seems that the excitation energy would vanish in this case. However, we note that the positive solution coincides with the unstable stationary point when this relation is satisfied, and therefore it does not correspond to a real spectrum. For finite field frequency, the coupling between two spectrums becomes much stronger as the coupling approaches the critical value. By considering this coupling, the eigenstates of the high-energy sector are given by the superposition of two different spectrums. Therefore the excitation energy of the high-energy spectrum does not become zero even in the vicinity of the critical coupling.

From previous discussion we see that the excitation energy vanishes only at $\varepsilon' = 0$ and $\kappa = 1$, demonstrating the existence of the QPT. We now discuss the critical behavior of the system as bias tends to zero. For $\kappa = 1$, the excitation energy vanishes as $\varepsilon' \to 0$ from either direction. In the $\varepsilon' \to 0^+$ limit, the Eq. (9) can be reduced to

$$\theta = -(2\varepsilon')^{\frac{1}{3}}. \tag{15}$$

Then the excitation energy $\omega'$ can be shown to vanish as

$$\omega' \sim \omega \varepsilon'^{\frac{1}{3}}. \tag{16}$$

Meanwhile, we identify the oscillator variance $\Delta(a + a^+)$ as the characteristic length scale. Our later calculations will show that this length diverges as $|\varepsilon'|^{-1/6}$, from which we find that $z = 2$ and $\nu = 1/6$. Here $\nu$ is the critical exponent and $z$ is the dynamic critical exponent.

**The role of A-square term**

In the minimal-coupling Hamiltonian, a term containing the square of the vector potential $\hat{A}^2$ is present. Although this term can be neglected in most cases, it can prevent the SPT in cavity QED systems [25-28]. We now consider the critical behavior of the biased Dicke model which contains the $\hat{A}^2$ term. With the $\hat{A}^2$ term included, an additional term $\kappa_0(a^+ + a)^2$ is added in the biased Dicke Hamiltonian. The parameter $\lambda$ and $\kappa_0$ are not independent of each other, and are related by $\kappa_0 = \alpha_0 N\lambda^2/\Omega$. Here $\alpha_0$ is an independent parameter decided by the field and the atoms. Then the Eq. (9) should be modified as

$$\left(\frac{\kappa^2}{1+\alpha_0\kappa^2} - \frac{1}{\cos\theta}\right)\sin\theta = \varepsilon' \quad (17)$$

The left hand side of Eq. (17) is a monotonically decreasing function for arbitrary coupling $\kappa$ since $\alpha_0 > 1$ is always satisfied according to the Thomas-Reiche-Kuhn sum rule. Thus the critical condition cannot be reached if the $\hat{A}^2$ term is not neglected, which is known as the no-go theorem[25].

The effective low-energy Hamiltonian can also be calculated by repeating the same step in the previous section, which reads

$$H_{CO} = \omega a^+ a + (N\beta\lambda\cos\theta + \kappa_0)(a^+ + a)^2 + C_1. \quad (18)$$

Here $C_1$ is a constant we do not care about. Then the excitation energy is given as

$$\omega' = \omega\sqrt{1 - \frac{4N\lambda^2\cos^2\theta}{\omega\Omega'} + \frac{4\kappa_0}{\omega}} = \omega\sqrt{1 - \frac{\Omega}{\Omega'}\kappa^2\cos^2\theta + \alpha_0\kappa^2}. \quad (19)$$

We note that the excitation energy can never vanish for arbitrary $\alpha_0$ provided $\varepsilon' \neq 0$. Noticing that parameter $\theta$ here is the negative solution to the Eq. (9) if we replaced $\kappa$ with an effective coupling constant $\kappa/\sqrt{1+\alpha_0\kappa^2}$. Since the excitation energy of the biased Dicke model remains finite for any coupling, we have

$$1 - \frac{\Omega}{\Omega'}\frac{\kappa^2}{1+\alpha_0\kappa^2}\cos^2\theta > 0. \quad (20)$$

With this inequility we find $\omega'/\omega > \sqrt{1 - (1+\alpha_0\kappa^2) + \alpha_0\kappa^2} = 0$, which indicates that in the CO limit the SPT cannot occur in the finite biased case regardless of the

$\widehat{A}^2$ term.

**Ground state wave function and its squeezing and entanglement**

We now consider the ground state wave function of the system in CO limit. In contrast to the doubly degenerate ground state in the original Dicke model, for finite bias the ground state remains nondegenerate even above the critical coupling.

According to our previous discussion, the ground state wave function in the CO limit is given as

$$|\psi(\theta_-)\rangle = e^{\frac{1}{2}i\theta_- J_y} D(\alpha) U(\beta,\gamma) S(\xi) |0\rangle |j,-j\rangle, \qquad (21)$$

where $D(\alpha)$ is the displacement operator. In this expression the operator $\exp(i\theta_- J_y/2) D(\alpha)$ comes from the mean-field results, $U(\beta,\gamma)$ is corresponded to the spin-oscillator correlation, while the operator $S(\xi)$ squeezes the bosonic field and produces the field fluctuations. It is interesting that no operator except $U(\beta,\gamma)$ in this expression is responsible for the spin fluctuations. Since the operator $U(\beta,\gamma)$ corresponds to the quantum corrections for the finite $\omega$ case, we can expect that the spin fluctuations vanish in the CO limit. In fact, the spin variance $\Delta J_x = \langle J_x^2 \rangle - \langle J_x \rangle^2$ for small $\omega/\Omega$ is of order $O(\beta^2)$, or equivalently $O(\omega^2/\Omega^2)$. The vanishing of $\Delta J_x$ as $\omega/\Omega \to 0$ shows that the spin fluctuations are strongly suppressed in the CO limit.

In order to discuss the fluctuations and squeezing of the field, we use the variance $(\Delta x)^2$ and $(\Delta p)^2$ of the field position operator $x = (a^+ + a)/\sqrt{2\omega}$ and the momentum operator $p = i(a^+ - a)\sqrt{\omega/2}$. Using the results from the Eq. (21) the variances are obtain as

$$(\Delta x)^2 = \frac{1}{2\omega} e^{-2\xi}, \qquad (\Delta p)^2 = \frac{\omega}{2} e^{2\xi} + O(\beta^2). \qquad (22)$$

Thus, we see that the field becomes a squeezed coherent state $|\alpha,\xi\rangle$ in the CO limit. In Fig. 5 (left panel) we plot the parameter $\xi$ as a function of the rescaled coupling constant $\kappa$. The parameter $\xi$ diverges at the QPT point ($\varepsilon' = 0$, $\kappa = 1$), which shows that the momentum variance is being strongly squeezed in the vicinity of the QPT point, and $(\Delta x)^2$ becomes strongly antisqueezed. This result is different from the CS limit case, where only a slight squeezing of the momentum variance presents as the system approaches the QPT point [5]. The field fluctuations, which can be described by $(\Delta x)^2$, diverge only at the QPT point. Near the QPT point, the parameter $\xi$ behaves as $e^{2\xi} \propto (\varepsilon')^{-1/3}$. Therefore both the fluctuations and the squeezing drop rapidly as the rescaled bias $\varepsilon'$ is increased from zero. This is agreed with our previous results, where the system experiences QPT only at the point $\varepsilon' = 0$ and $\kappa = 1$.

In addition to the fluctuations and the squeezing, we now study the spin-field entanglement. Here we use the entropy $S = -\text{Tr}(\rho \ln \rho)$ to quantify the entanglement, which is calculated with the reduced spin density matrix $\rho$. For $\varepsilon' = 0$ case, the $\mathbb{Z}_2$ symmetry in the original Dicke model will leads to symmetrized ground-state wave function $|\psi\rangle = (|\psi(\theta_-)\rangle \pm |\psi(\theta_+)\rangle)/\sqrt{2}$ when the system is in the superradiant phase. Thus the ground states obtain non-zero entanglement for $\kappa > 1$ in the CO limit. This entanglement will be destroyed in the present of finite bias due to the absence of $\mathbb{Z}_2$ symmetry and the ground state becomes nondegenerate. For $\varepsilon' \neq 0$ case, the ground state wave function can be written as $|\psi(\theta_-)\rangle = U(\beta,\gamma)|\alpha,\xi\rangle|\theta_-\rangle$. It is obvious that the entanglement is created completely by the operator $U(\beta,\gamma)$. Based on our previous discussion, the parameters $\beta$ and $\gamma$ become zero as $\omega/\Omega \to 0$, which indicates that the spin-field entanglement vanish in the CO limit. In Fig. 5 (right panel) we plot the entropy as a function of the rescaled coupling constant $\kappa$ for different $\omega/\Omega$. The vanishing of entropy as $\omega/\Omega \to 0$ shows the suppression of entanglement in the CO limit.

Different from the CO limit, the spin fluctuations are not suppressed in the CS limit. The effective model for the CS limit can be obtained by performing the Holstein-Primakoff transformation

$$J_+ = 2b^+\sqrt{2j - b^+b}, \quad J_- = 2\sqrt{2j - b^+b}\,b,$$
$$J_z = 2(b^+b - j), \qquad (23)$$

which is given as [22]

$$H_{CS} = \omega a^+a + \Omega' b^+b + \sqrt{N}\lambda\cos\theta(b^+ + b)(a^+ + a) - N\Omega'. \qquad (24)$$

This Hamiltonian is bilinear in the bosonic operators and can be simply diagonalized by the Bogoliubov transformation. Here we use the position operator and the momentum operator of bosonic mode $b$ in Eq. (24) to describe the squeezing in the spin. Then the variances in the ground state can be calculate, which are given as

$$(\Delta x_a)^2 = \frac{1}{2\omega_-}\cos^2\sigma + \frac{1}{2\omega_+}\sin^2\sigma, \quad (\Delta x_b)^2 = \frac{1}{2\omega_-}\sin^2\sigma + \frac{1}{2\omega_+}\cos^2\sigma,$$

$$(\Delta p_a)^2 = \frac{\omega_-}{2}\cos^2\sigma + \frac{\omega_+}{2}\sin^2\sigma, \quad (\Delta p_b)^2 = \frac{\omega_-}{2}\sin^2\sigma + \frac{\omega_+}{2}\cos^2\sigma, \quad (25)$$

where the subscripts denote the bosonic modes $a$ and $b$ respectively. In Fig. 6, we show the analytical values of these variances as a function of κ for different biases. We see that a finite bias will diminish the sharp increase of $(\Delta x_a)^2$ and $(\Delta x_b)^2$ as κ approaches QPT point in the zero biased case. The slight squeezing of $(\Delta p_a)^2$ and $(\Delta p_b)^2$ is also weakened when bias is increased. However, the bias has little effect on the squeezing as the coupling constant κ is far from the QPT point, and the variances stay largely constant for different bias.

**Conclusion**

We have analyzed the properties of the biased Dicke model in the CO limit. For finite bias, the mean-field results show that the ground state remains nondegenerate even above the critical coupling. The low-energy effective Hamiltonians are found for the CO limit. We find that the low-energy spectrum turns into two equally spaced spectrums when the coupling is above the critical value, and the energy gap between these two spectrums is much larger than the excitation energy, thus we can reasonably regard the lower one as the low-energy spectrum. For finite bias, the excitation energy remains finite for arbitrary coupling. This indicates that the second order phase transition is avoided for finite bias in the CO limit.

We then discuss the excitation energy of the biased Dicke model in the presence of the $\widehat{A}^2$ term. The resulted effective low-energy Hamiltonian is given and we show that, the SPT cannot occur for the finite biased case regardless of the $\widehat{A}^2$ term. The results also demonstrate that the $\widehat{A}^2$ term prevents the system from reaching the critical coupling, thus can be compensated by other interaction terms. However, a finite bias will always break the $\mathbb{Z}_2$ symmetry and destroy the superradiant phase completely. Thus the SPT is hard to recover unless we manage to restore the symmetry.

We have also calculated the ground state and discuss its squeezing in both limits. The results show that the bias term will suppress the squeezing strongly in the vicinity of the QPT point, and has little effect as the coupling is far from the QPT point. Moreover, the squeezing properties are quite different in two classical limits. In the CO limit, the momentum variance of the field is being strongly squeezed in the vicinity of the QPT point, in comparison with the slight squeezing of the momentum variance in the CS limit.

The entanglement of the spins and the field will behave dramatically different for finite bias. Instead of remaining a finite value [10], the entanglement vanishes as we approach the CO limit. Thus the entanglement of the spins and the field will be suppressed in the presence of the bias term.

**Figure captions**

Figure 1: Graphical solutions of Eq. (9) for different values of coupling $\kappa$. The $a, b, c$ lines represent the left side in Eq. (9), while the horizontal line is the right side in the equation.

Figure 2: The rescaled energy levels of the first 30 excited states relative to the ground state energy as functions of the coupling strength $\kappa$. Here we take $\Omega/\omega = 60$, $\varepsilon' = 0.11$ with $N = 5$.

Figure 3: The rescaled excitation energy as functions of the coupling strength $\kappa$. The dashed line represents the analytical result in Eq. (14), while the $a, b, c$ lines are the excitation energies of the first excited state relative to the ground state. Here we take $\varepsilon' = 0.11$ with $N = 5$.

Figure 4: The excitation energies as functions of the coupling strength $\kappa$ and the bias $\varepsilon'$ in the

CO limit.

Figure 5: (Left panel) The squeezing parameter ξ of the ground state as a function of coupling κ for different biases in the CO limit. (Right panel) The entanglement entropy S of the ground state as a function of coupling κ for different $\Omega/\omega$. Both panels start from the curve for $N = 5$.

Figure 6: The squeezing variances of the field (left panel) and the spin (right panel) of the ground state for different biased cases in the CS limit. Here solid lines denote the variances of the position operators, whereas dashed lines correspond to the variances of the momentum operators. The Hamiltonian is on scaled resonance: $\omega = \Omega = 1$.

**Acknowledgments**

This work is supported by the National Natural Science Foundation of China (Grant No. 11574022 and 11174024) and the Open Fund of IPOC (BUPT) grants Nos. IPOC2013B007, also supported by the Open Project Program of State Key Laboratory of Low-Dimensional Quantum Physics (Tsinghua University) grants Nos. KF201407, and Beijing Higher Education (Young Elite Teacher Project) YETP 1141.

**Author contributions**

G.Z. conceived and designed the research. G.Z. and H.Z. performed analysis and wrote the manuscript. H.Z. prepared all the figures. H.F. reviewed the paper and gave some valuable suggestion.

**Additional information**
Competing financial interests: The authors declare no competing financial interests.

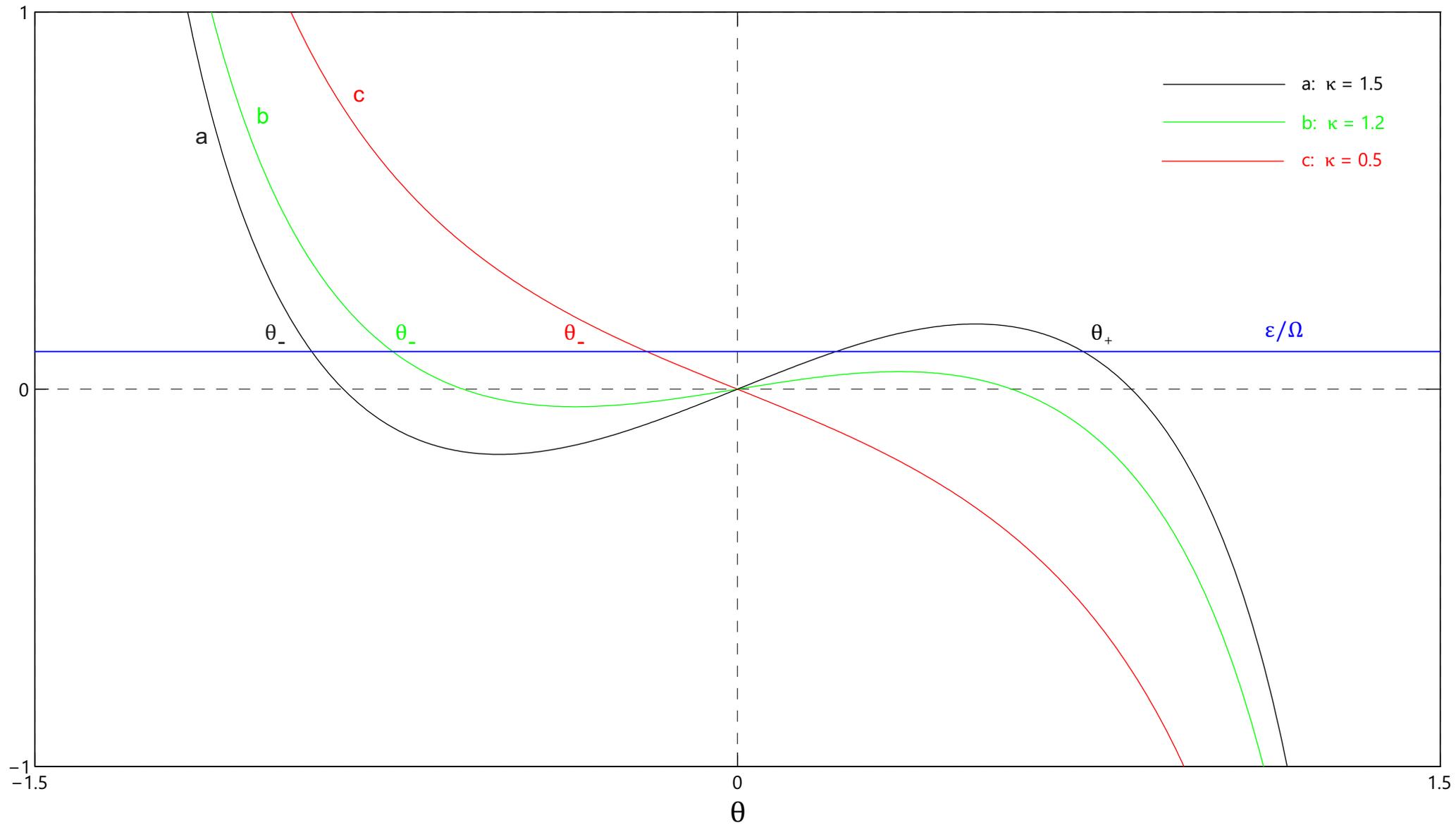

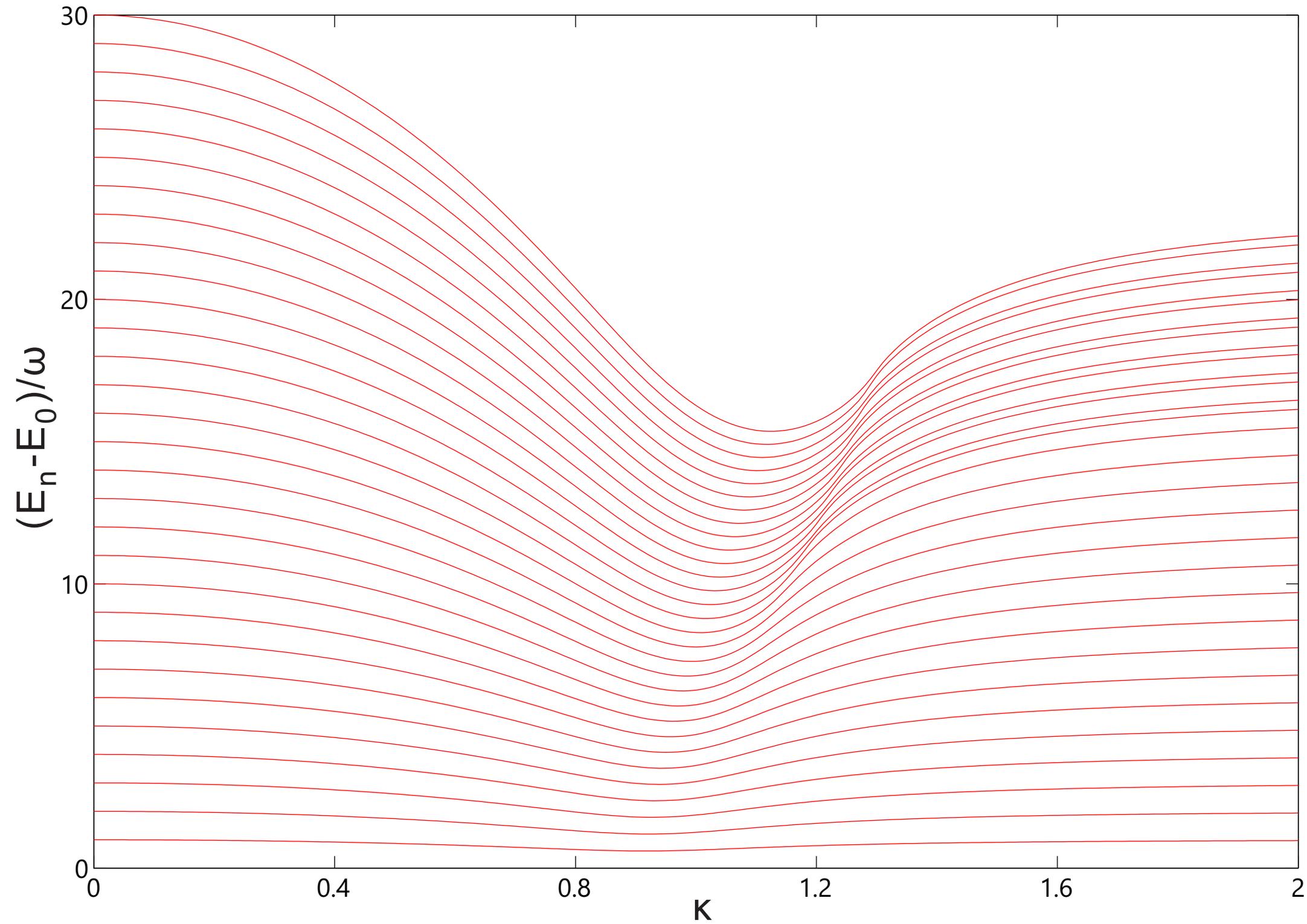

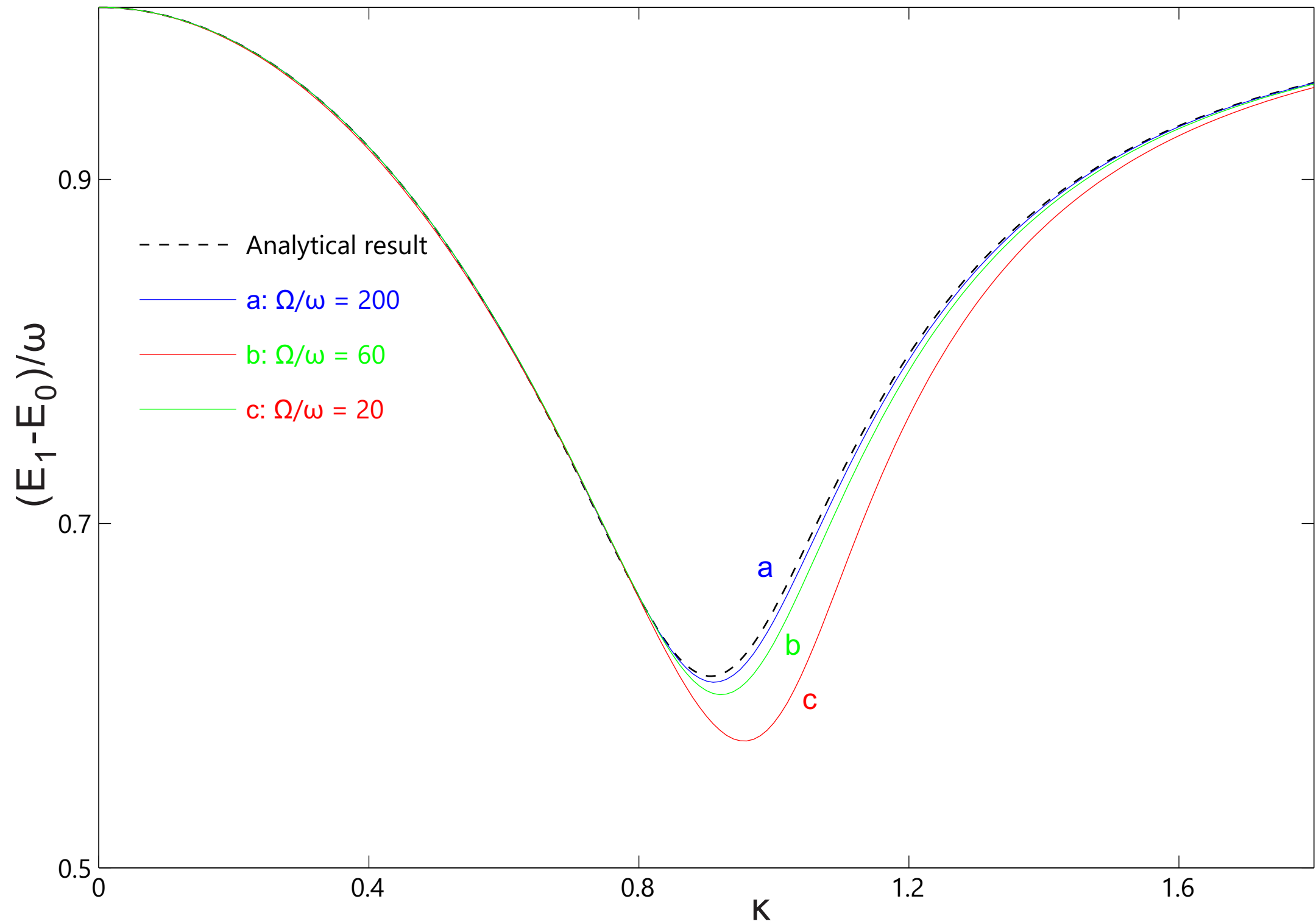

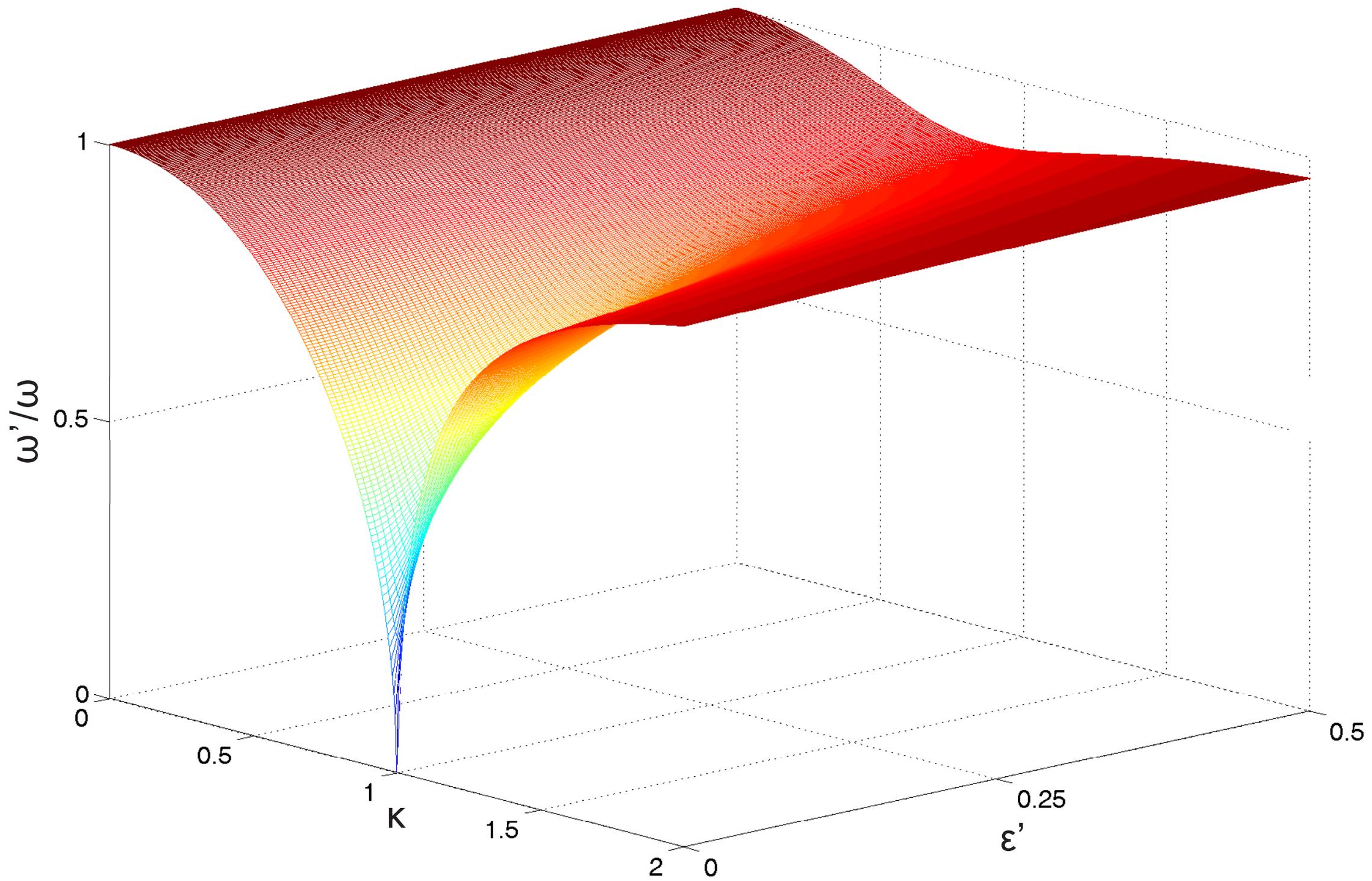

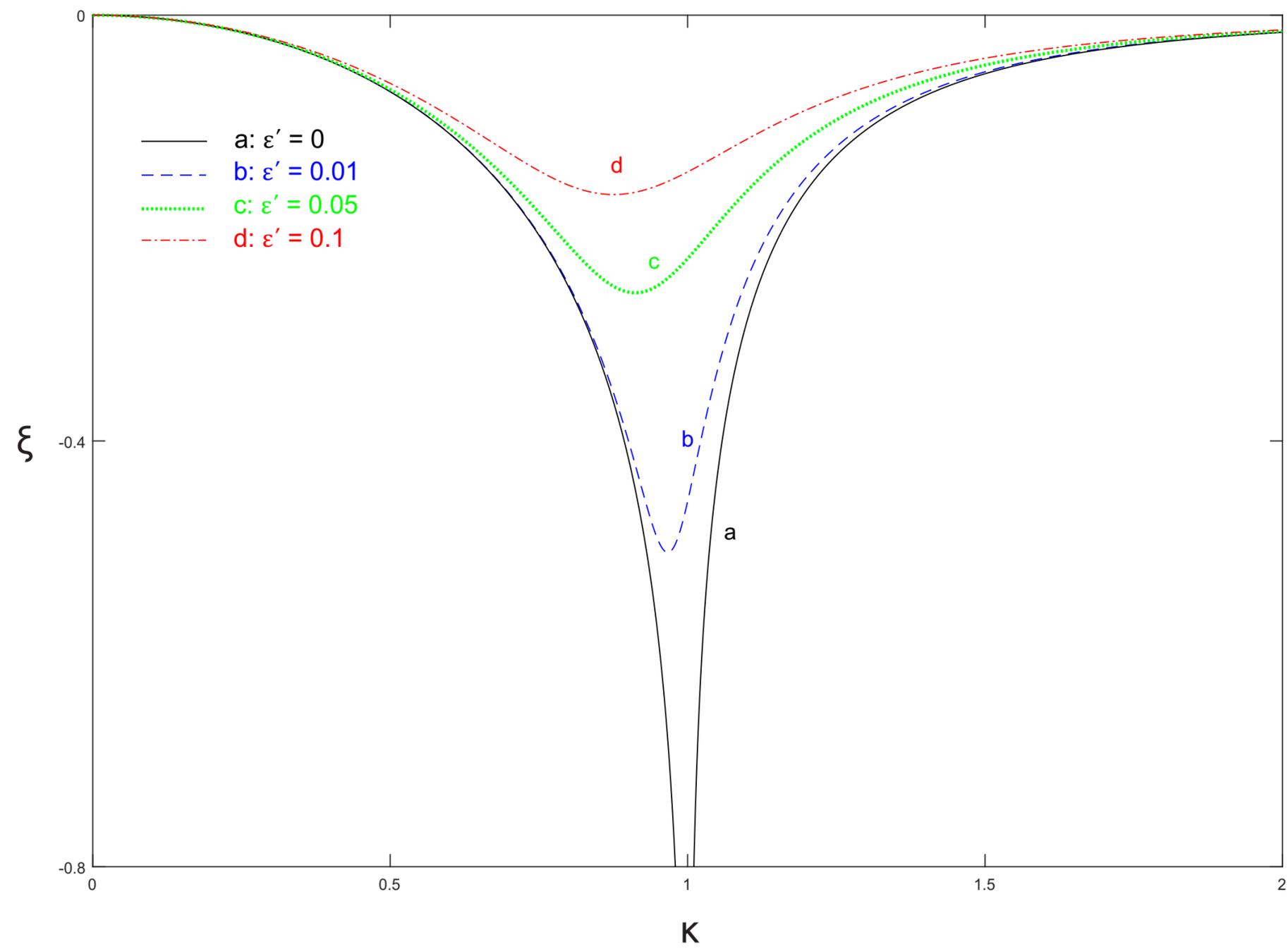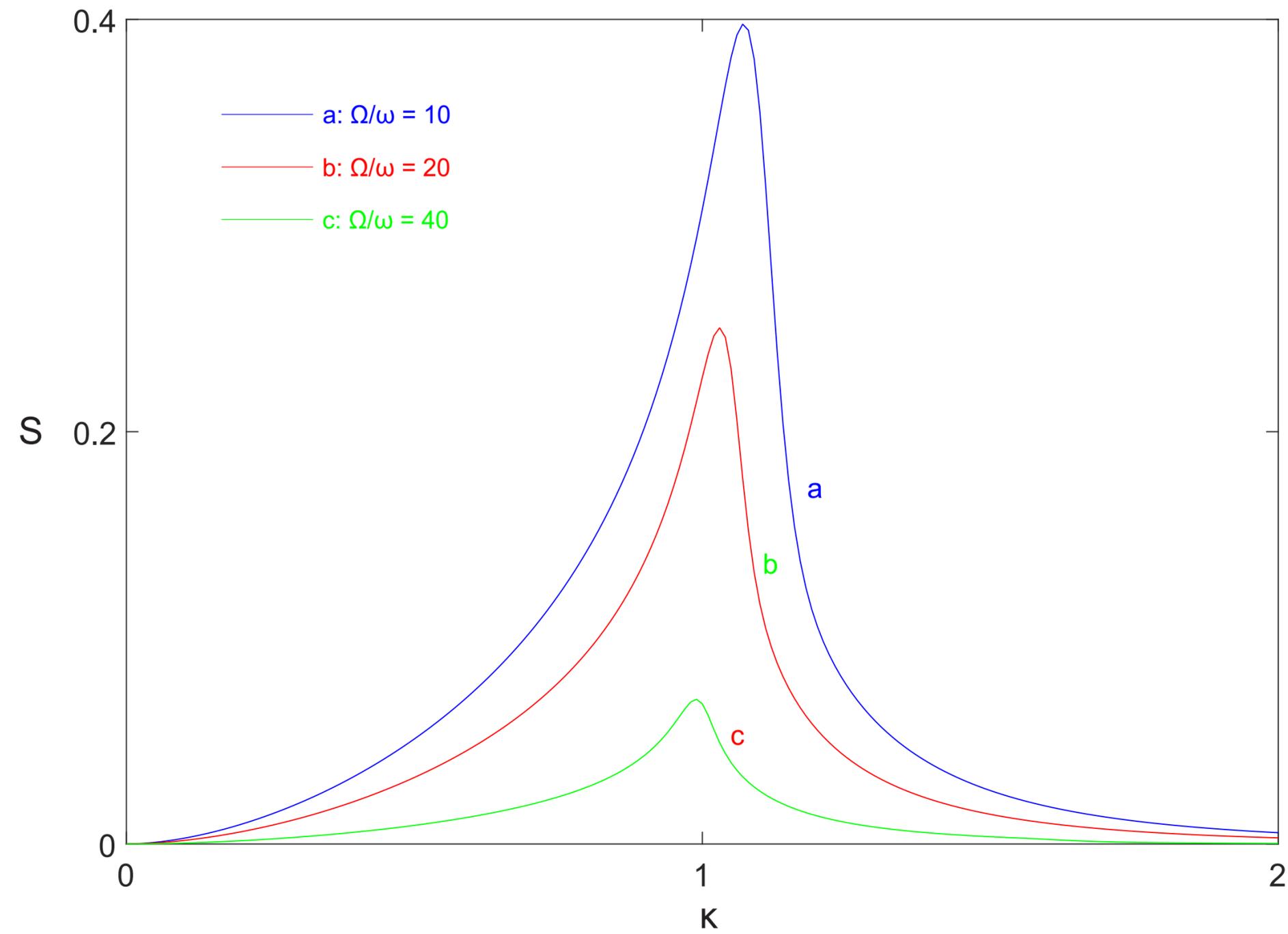

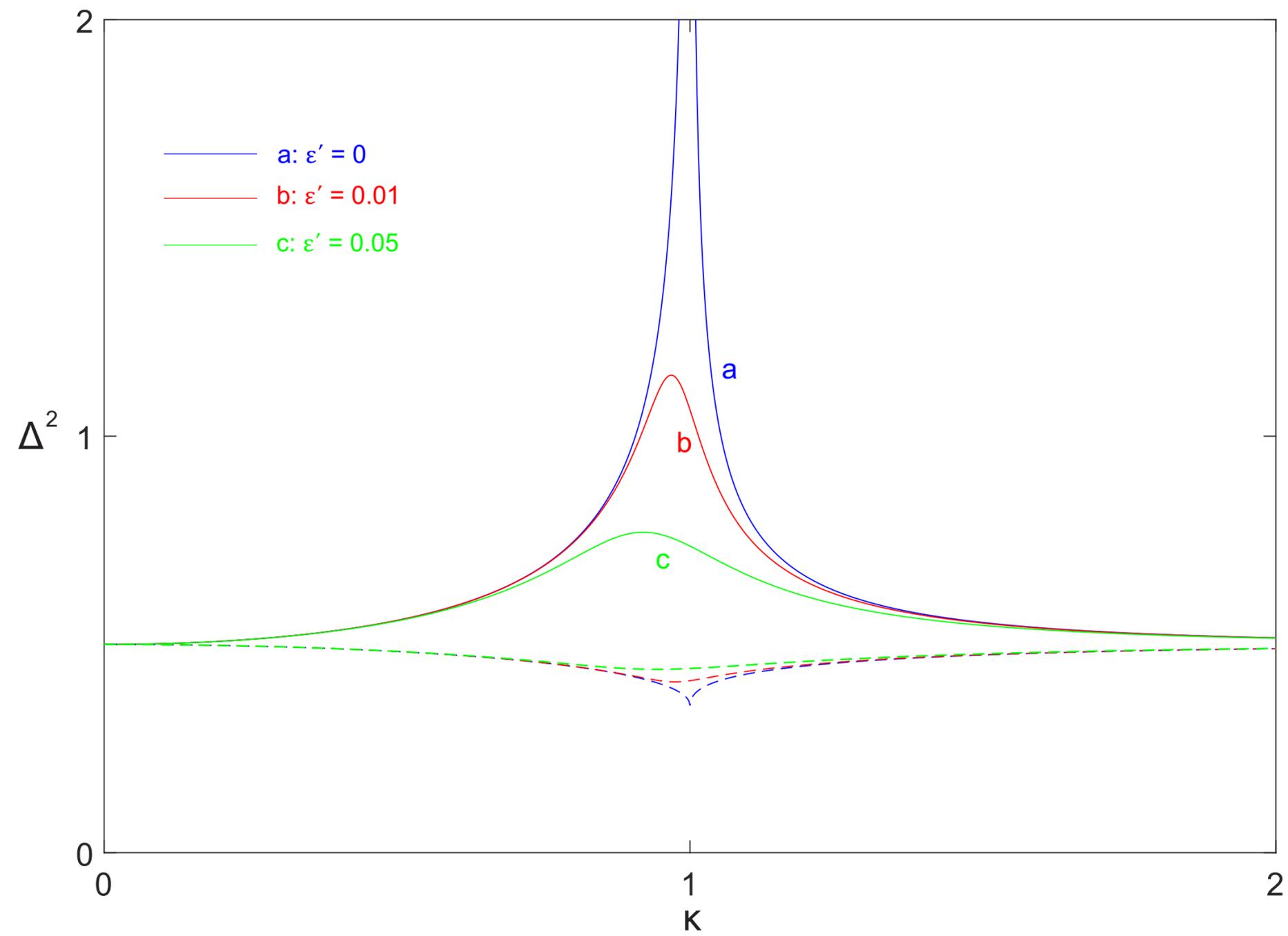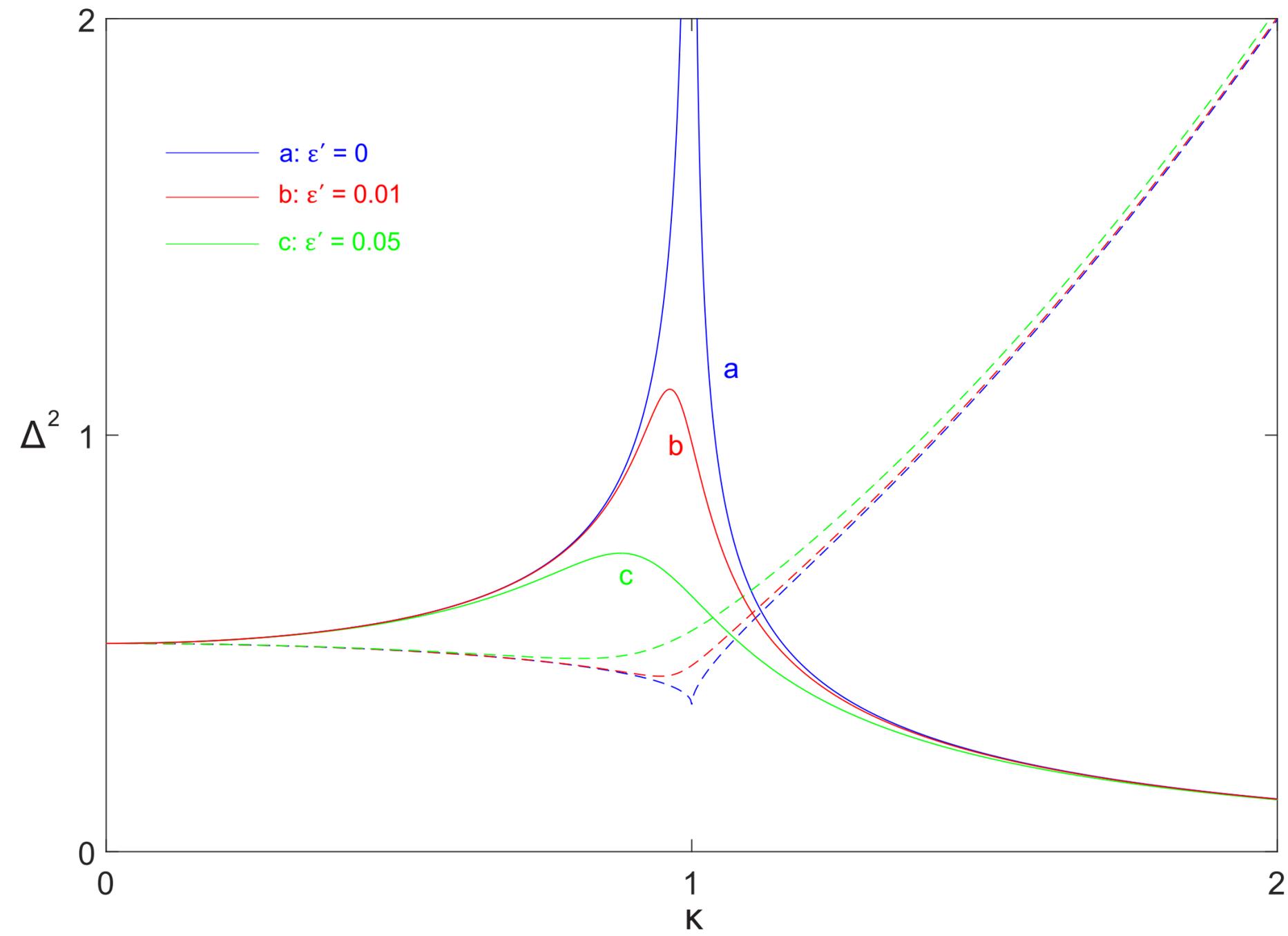